*Chapter*

# LARGE-VOLUME OPTICAL COHERENCE TOMOGRAPHY WITH REAL-TIME CORRECTION OF GEOMETRIC DISTORTION ARTIFACTS

*Sam Van der Jeught,[1, *] Jan A.N. Buytaert[1], Adrian Bradu[2], Adrian Gh. Podoleanu[2] and Joris J.J. Dirckx[1]*

[1]Laboratory of Biomedical Physics, University of Antwerp, Groenenborgerlaan 171 B-2020 Antwerp, Belgium
[2]Applied Optics Group, School of Physical Sciences, University of Kent, CT2 7NH Canterbury, United Kingdom

## ABSTRACT

Large-volume optical coherence tomography (OCT)-setups employ scanning mirrors and suffer from non-linear geometric distortion artifacts in which the degree of distortion is determined by the maximum angles over which the mirrors rotate. In this chapter, we describe a straightforward approach to correct for these distortion artifacts, creating an alternative to previously reported ray-tracing schemes that are unable to apply these corrections in real-time. By implementing the proposed 3D recalibration algorithm on the graphics card of a standard computer, this feature can be applied in real-time. We validate the accuracy of the technique using OCT measurements of a highly curved object within a large imaging volume of 12.35 x 10.13 x 2.36 mm³. The resulting 3D object shape measurements are compared against high-resolution and aberration-free optical profilometry measurements. Maintaining an optical resolution of <10μm within the sample, both axially and transversally, we realized a real-time, high-resolution, large-volume OCT imaging system, capable of producing distortion corrected wide-field OCT data with a geometric surface shape accuracy of <15μm.

* Corresponding Author address
  Email: sam.vanderjeught@ua.ac.be

## INTRODUCTION

Optical coherence tomography (OCT) is a noncontact and noninvasive interferometric technique allowing micrometer resolution imaging of sample substructures (Huang et al. 1991). Broadly, OCT can be subdivided into two categories: time domain (TD)-OCT and spectral domain (SD)-OCT (Schmitt 1999, Fercher et al. 2003). In this chapter, a spectrometer based (SB) spectral domain OCT technique is used. This technique employs a Michelson interferometer equipped with a broadband light source, illuminating the observed object in the sample arm and the reflective mirror in the reference arm. The reflected beams are guided towards an optical beam splitter, where they recombine before their interference signal is captured by a high-resolution digital camera. A succession of axial depth scans (or A-scans) are used to form a two-dimensional cross-section image (or B-scan) of the object. SB-OCT holds the reference mirror at a fixed position and treats the resulting interference signal as a modulation in wavenumber $k$. Hence, SB-OCT can obtain a full A-scan of the sample by processing the spectrum data, in opposition to TD-OCT where in one measurement step only the signal from a particular depth is obtained. This leads to the higher speed capabilities of SB-OCT in comparison with TD-OCT (Leitgeb et al. 2003) in addition to superiority in sensitivity conferred by spectral processing (Choma et al. 2003). The introduction of high-speed OCT imaging has been of particular interest to the medical community as it allows for *in vivo* three-dimensional volumetric reconstructions of sample microstructures (Kraus et al. 2012, Kerbage et al. 2007, Yasuno et al. 2005, Miura et al. 2007). Two-dimensional images can be insufficient to recognize features of interest, such as blood vessels or tissue boundaries, when the imaging plane in which the individual image is taken only contains one cross-section of the larger feature. Computer-based volume rendering grants the operator considerable diagnostic advantages. A full display of the entire volume of the imaged object is made accessible to the operator, with options in terms of selecting the perspective of viewing and slicing the volume (Drexler & Fujimoto 2008).

The capability of displaying volumetric information has stimulated numerous applications in various biomedical fields including cardiology (Raffel 2008), ophthalmology (Radhakrishnan et al. 2001, Wojtkowski et al. 2005, Nassif et al. 2004), dentistry (Amaechi et al. 2001) and larynx and cochlea imaging (Bibas et al. 2004, Sepehr et al. 2008). Besides real-time monitoring benefits, high-speed OCT has also enabled larger wide-field scans to be made without mechanically moving the sample or the imaging lens through the use of scanning mirrors. However, increasing the field of view also increases the angle over which the scanning mirrors need to rotate. As this angle increases, the amount of geometric distortion imposed onto the resulting images, often referred to as *fan distortion,* increases as well. To compensate for this effect, numerous authors have reported the implementation of ray-tracing algorithms in their large-field

OCT scanning setups (Westphal et al. 2002, Zhao et al. 2010, Ortiz et al. 2009 & 2012, McNabb et al. 2012), programmatically applying Fermat's principle. Besides correcting for fan and lens distortion artifacts, ray-tracing also takes the relatively less significant refraction artifacts into account that may occur due to the scanning beam traversing media of different refractive indices. Ray-tracing, however, is fundamentally unable to apply these corrections in real-time, as it requires prior knowledge of the sample microstructure and typically demands high processing power. Though many applications of large-field OCT imaging, such as the ones listed above, would benefit greatly from having geometric fan distortion artifacts corrected for in real-time, operators currently rely on post-processing to calculate the actual shape of the observed sample.

In the following, we describe a 3D recalibration process that compensates for these distortion artifacts by applying successive coordinate transformations to the acquired data block. FD-OCT measurements of a highly curved object are compared with high-resolution optical Moiré profilometry results, both before and after applying the distortion correction algorithm, to evaluate the quality of the proposed reconstruction scheme. Furthermore, we investigate the real-time feasibility of the presented correction algorithm by implementing the necessary calculations in parallel on a graphics processing unit (GPU) of a standard PC. As we assume there to be no *a priori* knowledge of the observed sample, the presented method does not account for refractive errors. However, given the fact that refraction only occurs after the scanning beam has crossed the sample surface, the observed surface shape itself can still be visualized correctly in real-time. This feature could prove to be useful for the earlier mentioned 3D-OCT applications and especially essential for various applications that monitor topographic information of biomedical tissues such as corneal topography (Bechmann et al. 2001) and tympano-topography (von Unge et al. 1993), both in research and medical diagnostics. Real-time application of such a capability may provide the operator with considerable diagnostic and orientational advantages before any additional refraction errors are eliminated in post-processing.

## EXPERIMENTAL SETUP

A schematic diagram of the employed FD-OCT system is shown in fig. 1. A broadband light source, based on an Ytterbium-doped fiber (Multiwave Photonics, Porto, Portugal) similar to the one described by Trifanov et al. (2011) with a central wavelength at $\lambda = 1050$ nm and bandwidth $\Delta\lambda = 70$ nm was used. The optical signal originating from the light source is divided into reference and sample arm by a directional coupler (DC, splitting ratio 80/20). In the sample arm, the beam is collimated by the microscope objective MO1 and diverted towards the sample via the scanning unit $GXY$ and the microscope objective achromatic lens L1 with working distance of 20 mm. The scanning unit $GXY$ consists of two galvo-scanning mirrors, $S_x$ and $S_y$, whose respective axes of rotation are perpendicular to each other. Backscattered light from the sample is collected and guided towards DC where it interferes with the light returning from the reference arm.

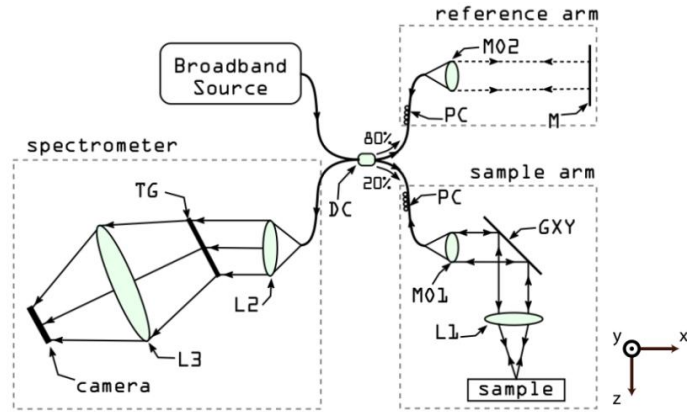

Figure 1: Schematic of the employed FD-OCT imaging system. Broadband light source, MO1-2: microscope objectives, L1-3: achromatic lenses, DC: directional coupler, PC: polarization controller, GXY: orthogonally placed galvo-scanning heads, TG: transmissive diffraction grating.

The interferometric signal is conveyed towards a spectrometer which comprises two achromatic lenses (L2-3), a transmission grating (TG, 1450 lpm, Wasatch Photonics, Logan, Utah) and an InGaAs line-scan camera (SU-LDH, Goodrich-SUI, Princeton, New Jersey), operating at 46.9 kHz. Data from the linear CCD is transferred to a PC via a Cameralink cable and a high speed PCI Express frame grabber (NI-1427 National Instruments, Austin, Texas). Each of the two rotating mirrors in the galvo-scanning unit $GXY$ is provided with a combination of an input voltage source and a function generator. In order to ensure that the mirrors remain rigid even when subjected to large accelerations, the input voltage waves are chosen to be symmetric triangles. As the incoming light beam reaches the mirrors sequentially, both mirrors have different optical distances to the sample, leading to different scanning angles to cover the same distance in object space along the $X$-and $Y$-axis. In our setup, a maximum voltage ramp of 7.5 V peak-to-peak is applied on the mirror which is positioned the furthest from the object surface ($Y$-scanner), and a maximum voltage ramp of 19.3 V peak-to-peak is applied on the mirror which is positioned the closest to the object surface ($X$-scanner). The maximum line-scan rate of the employed spectrometer limits the maximum frequency at which the respective mirrors may rotate. While the frequency $f_x$ of the $X$-scanner determines the width of the B-scan OCT image, the frequency $f_y$ of the $Y$-scanner determines the number of B-scans per volume. Because of the symmetric shape of the applied voltage waves, the maximum frequencies of the respective scanners must be divided by two in order to avoid mirror imaging. In general, these restraints yield the following expressions for $f_x$ and $f_y$:

$$\begin{cases} f_x \leq \dfrac{f_{cam}}{2N_x} \\ f_y \leq \dfrac{f_x}{2N_y} \end{cases} \quad (1)$$

with $N_x$ the number of pixels in the X-direction of the B-Scan, $N_y$ the number of B-Scans per 3D volume and $f_{cam}$ the maximum line-scan rate of the camera. In the current configuration, $f_{cam}$ is limited to and set at 46992 lines/s, while $N_x$ and $N_y$ are fixed at 1024 pixels each. Employing these settings in our experimental setup limits $f_{x,MAX}$ to 22.9 Hz and $f_{y,MAX}$ to 0.011 Hz. In the following measurements, however, we will use 21.9 Hz and 0.009 Hz in the X-and Y-scanning directions, respectively, to apply a safety margin. These scanning frequencies, in combination with the above mentioned voltages lead respectively to a maximum imaging field of 12.35 mm by 10.13 mm at the working distance of the object lens L1. A full data-acquisition of a single 3D stack of 1024 B-scans of 1024 pixels wide and 1024 pixels deep takes around 100 seconds on average to record without image processing and 300 seconds on average including image processing. After removal of the identical information carried by the opposite sign of the spectrum, the effective pixel dimension of the data block in depth is halved to 512 pixels. The axial image size and resolution of the OCT-system are determined experimentally by imaging a flat mirror with the given setup. The axial OCT signal (A-scan) of this mirror forms a narrow peak, indicating the relative position in depth of the reflective mirror plane. By moving the mirror axially and by measuring the total distance this mirror travels within the A-scan, the axial image size is determined with the precision of the micrometer-resolution translation stage. The axial resolution of an OCT setup is given by the coherence length $l_c$ of the light source and is largely determined by the optical bandwidth $\Delta\lambda$:

$$l_c = \frac{2\ln(2)}{\pi n} \frac{\lambda^2}{\Delta\lambda} \quad (2)$$

where $\lambda$ represents the central wavelength of the optical source and $n$ the refractive index of the imaged medium. This theoretical approximation of the coherence length assumes a Gaussian distribution of the wavelengths emitted by the light source. However, in reality, this is most often not the case (Tripathi et al. 2002). Therefore, we determined the actual axial resolution by measuring the full width at half maximum (FWHM) of the A-scan signal peak. For our setup, this FWHM remains under 10 µm throughout the maximum sample depth of 2.36 mm. The transversal resolution and depth of focus of OCT-setups are decoupled from the axial resolution and are limited mainly by the focal spot size of the Gaussian profile light beam. The employed configuration maintains a lateral resolution of 10 μm across the sample with a depth of focus that covers at least the full axial scanning depth of 2.36 mm. Lateral field-of-view is measured in both $X$-and $Y$-direction by tracking a feature of the imaged object in the OCT image at the working distance of the imaging lens, and by moving the object laterally using a translation stage so that this feature crosses the entire image from one side to the other. Our experimental setup produces a lateral field-of-view in the $X$-and $Y$-direction of 12.35 mm and 10.13 mm, respectively.

## METHODS

### Index of refraction correction

In order to maintain accurate thickness information of the observed object, the difference in index of refraction between subsequent layers of the sample must be accounted for. Usually, this is done by simply dividing the full length of the imaged A-scan with the average index of refraction of the observed sample Kim et al. 2009), including the parts that represent air. Despite the fact that this approach does rescale the average thickness distribution of the sample correctly, it does not produce an accurate representation of the internal structure of the object. As the parts in air are rescaled with the same constant, their respective lengths influence the relative starting positions of the deeper sample layers where they should not. To this end, we algorithmically correct for the parts that have an index of refraction within the sample of $n > 1$ and maintain the respective lengths of the parts in air, as illustrated in fig. 2.

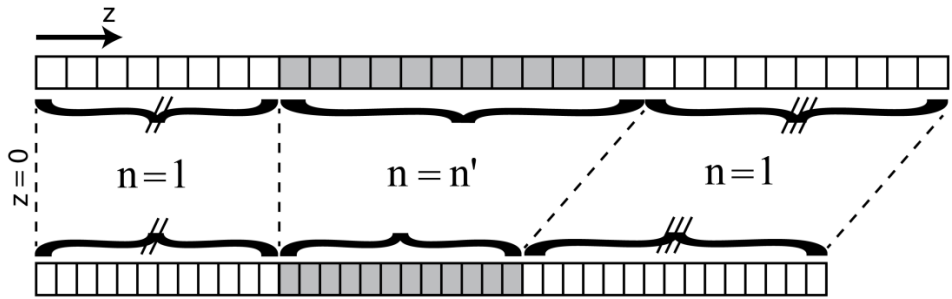

Figure 2: A-scan representing the OCT data of the observed sample (grey) with index of refraction $n = n'$ and the parts in air (white) with index of refraction $n = 1$, before (top) and after (bottom) index of refraction correction.

By doing so, both the thickness and shape information of the original sample will be preserved. Notice how rescaling the actual object data (grey pixels) causes the pixels to be repositioned more closely to each other. Instead of undersampling the OCT data by maintaining the same sampling distances as before, we opt to sample the full A-scan at a new sampling rate, determined by the repositioned object data. This causes the parts in air (white pixels) to be oversampled, but preserves all original OCT data in the relevant regions.

**Origin of geometric distortion**

The optical path length of the probing beam is determined by the relative positions of the scanning mirrors $S_x$ and $S_y$. When the rotational angles of these scanning units change, so does the optical path length of the light beam as it traverses the surface of the imaged sample, creating a dependency in depth of the OCT system on the rotational position of the scanning mirrors. This effect results in a geometric distortion artifact that causes a flat surface to appear curved in the recorded OCT image. Although this type of distortion may be negligible for imaging applications that require only small areas to be scanned under small rotational angles, it can no longer remain uncorrected for when performing larger field-of-view recordings. Such cases require accurate shape and distance measurements across the entire sample surface. An illustration of the standard 3D scanning geometry of the employed OCT setup is included in fig. 3, in which an enlarged detail of the scanning configuration at the end of the object arm is shown, performing the 3D recording of a flat mirror (FM). The probing beam crosses the collimation-condensing lens L1 at a location determined by the tilting angles of the two scanning mirrors $S_x$ and $S_y$, each independently capable of creating B-scans (labeled $B_x$ and $B_y$, respectively) in mutually perpendicular directions. By scanning the beam along the X-axis using the line-scanning mirror, whilst keeping the frame scanning mirror fixed, a B-scan cross section is produced in the $XZ$-plane with coordinate $Y$ determined by the tilt of the frame scanning mirror. Repeating the process for multiple adjacent $Y$-values, a stack of B-scan images is acquired. Typically, the line scanning mirror rotates at a higher frequency than the frame scanning mirror. Note the difference in curvature between the two transversal scans $B_x$ and $B_y$ of the same flat mirror. As the two scanning mirrors reflect the probing beam consecutively, their respective beam reflecting points each have a different optical path length to the surface of the sample. Normally, if the distance between the pivot of the galvoscanner and the lens is equal to the focal length of the lens, then the wavefront behind the lens should be flat. However, this distance cannot be adjusted precisely and therefore the wavefront behind the lens deviates from a flat surface. There is also a distance of 1 cm between the pivots of the two mirrors. This is rather large in comparison to the focal length of the lens, of 2.5 cm. This means that if the pivot of one galvoscanner is placed at f, then the pivot of the other galvoscanner is at a different distance than f and the wavefront cannot be flat for both. More sophisticated configurations consist in galvoscanners separated by telescopes in which case the wavefront can be flat for both. In this regard, when the OCT setup is configured similarly to the one depicted in fig. 3, the amount of angular distortion imposed onto $B_y$ by scanning mirror $S_y$ will be significantly less than that imposed onto $B_x$ by scanning mirror $S_x$ as the origin of distortion of $B_y$ lies further away from the object surface than that of $B_x$.

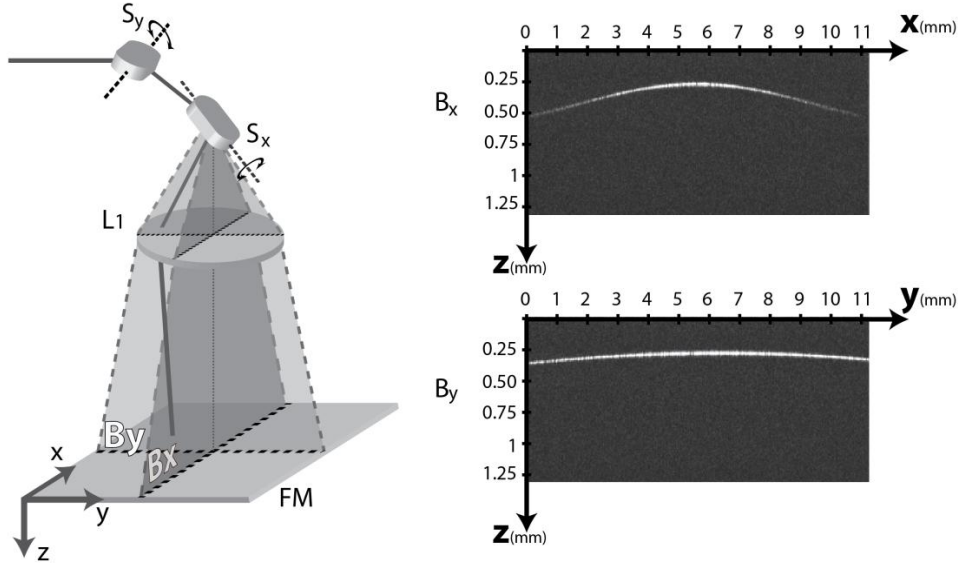

Figure 3: Enlarged detail of the scanning unit in the object arm in the OCT setup. Rotating one of the two scanning mirrors ($S_x$ or $S_y$) whilst keeping the other one fixed, produces a B-scan ($B_x$ or $B_y$) perpendicular to its respective axis of rotation.

### First order distortion correction in 3D

By scanning the mirror in the $X$-direction whilst keeping the $Y$-scanning mirror fixed, and vice versa, it can be considered that the geometric distortion effects imposed onto the OCT volume by the two galvoscanners are decoupled. This allows gauging the distortion characteristics of the two scanners separately and compensating for them sequentially. To calibrate the 3D stack of OCT data, a series of B-scans of a flat reflective mirror was obtained using the above described OCT setup. Each image was processed as illustrated in fig. 4. First, threshold filtering was applied to the greyscale OCT images (fig. 4A, 4B) of the observed mirror, to infer the topography of the surface. Next, the analytical equation of a circle was fitted to these data points using least-squares fitting (fig 4C, 4D). In this way, the perceived origin location of radial distortion can be determined for each B-scan. The virtual scanning radii $R_x$ and $R_y$ are defined as the shortest distances between the perceived point source of distortion and the surface of the observed flat mirror, in the respective scanning directions.

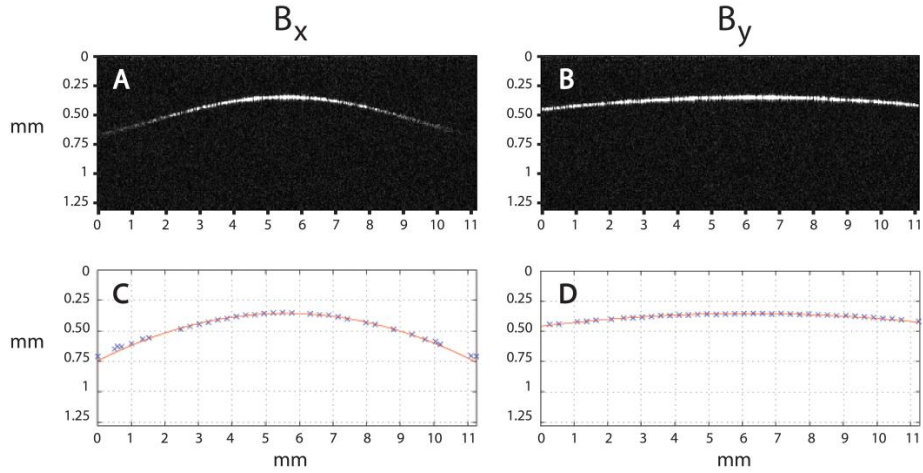

Figure 4: Calibration images are recorded to compensate for radial distortion artifacts caused by rotating scanning mirrors to scan the beam along the $X$- direction (A, C) and along the $Y$- direction (B, D). After thresholding and least-squares fitting of an analytical circle, the radii $R_x$ and $R_y$ of the respective origins of distortions can be determined.

Slight deviations from a perfect circular fit can be noticed near the far edges of the object B-scan. As the scanning beams are focused onto the observed object by imaging lens L1, higher order distortion effects are superimposed on the first order radial distortion artifacts in the image. These effects, however, do not preclude the overall radial nature of the distortion effect.

To cover the entire observed surface, B-scans of the flat mirror were recorded at 100 different angles of the fixed mirror, and in both scanning directions. This process is repeated at several depth locations within the observed volume by translating the calibration mirror axially until the full axial range of the imaging volume has been crossed. By interpolating between these collections of acquired radii, proper three-dimensional recalibration can be achieved after applying successive coordinate transformations to the acquired OCT data, as shown schematically in fig. 5. In practice, both distortion effects are imposed onto the imaged volume simultaneously. However, when one of the scanning mirrors is rotating whilst the other one is kept fixed, and vice versa, the scanning directions can be considered decoupled and the order in which the data block of OCT images can be corrected becomes interchangeable. The B-scans that comprise the three-dimensional OCT volume are stacked in system memory in the form of an isometrically sampled data cube $D(x, y, z)$, containing the recorded intensity values $I_{x,y,z}$ at each voxel coordinate location $(x, y, z) \in \mathbb{N}^3$. However, as shown earlier, this Cartesian spacing of acquired intensity values does not correlate with the geometric shape in which the B-scans are actually recorded. As the scanning mirrors $S_x$ and $S_y$ arch the OCT probing beam path, the resulting cross-sections become curved, as illustrated in part I of fig. 5 by the bent black line representing the recorded flat mirror.

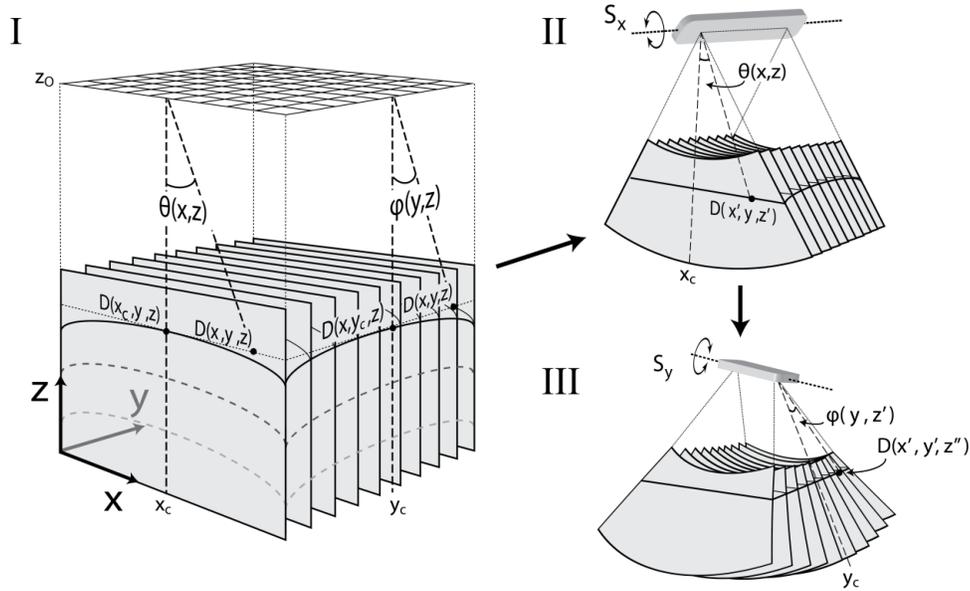

Figure 5: Calibration procedure consisting of successive coordinate transformations that effectively compensate for any radial distortion artifacts induced by rotating scanning mirrors $S_x$ and $S_y$. The curvature depends on the position of the lens between the scanner and the object, for simplicity this is not shown.

Both scanning directions introduce their own curvature artifacts, defined by the virtual radii $R_x$ and $R_y$, so in order to ensure proper geometric calibration, the coordinate locations of $D(x, y, z)$ need to be relocated in such a way that the imposed curvature artifacts are compensated for in both scanning dimensions. In the included example, the distortion effect caused by the rotation of $S_x$ will be corrected first, followed by correction of effects due to the rotation of $S_y$.

As a first step in the data calibration procedure, the 2D Cartesian coordinate points $(x, z)$ on every $B_x$-scan in the decoupled $Y$-direction are associated with their polar coordinate system equivalents $(R, \theta)$ originating from the virtual origin of distortion $(x_c, z_o)$. After having experimentally determined the full set of radii $R(x_c, z)$, measured at the central column $(x = x_c, z)$ of each $B_x$-scan and for every $z$, the angular component $\theta(x, z)$ of the 2D polar coordinates can be calculated for all $B_x$-scans using basic geometry:

$$\forall z \begin{cases} R(x_c, z) \quad experimentally\ determined \\ \forall x: \ \theta(x,z) = tan^{-1}\left(\dfrac{x - x_c}{R(x_c, z)}\right) \end{cases} \quad (3)$$

where $\theta(x,z)$ represents the angle between $\vec{R}(x,z)$ and the central column line $\vec{R}(x_c, z)$ of the currently processed $B_x$-scan. Note that the coordinate transformation procedure can be written independently of the decoupled scanning direction $S_y$, indicating that the origin of distortion of $S_x$ can temporarily be seen as a *line* source of distortion points, containing one origin point of distortion for every processed $B_x$-scan. Next, the same value of $R(x_c, z)$ is assigned to all coordinate points that have the same z-values, and this process is repeated for all $B_x$-scans. By maintaining the respective angles $\theta(x,z)$ of all coordinate points, the data block is reshaped so that the distortion effect caused by the first scanning direction is corrected for, as illustrated in part II of fig.5:

$$\forall x, z \begin{cases} R(x,z) := R(x_c, z) \\ \theta(x,z) = \theta(x,z) \end{cases} \quad (4)$$

This polar coordinate system can be reconverted to its Cartesian equivalent by applying the following transformation to all $(R, \theta)$-coordinate couples:

$$\forall x, z \begin{cases} x' = R(x_c, z) \sin(\theta(x,z)) \\ z' = R(x_c, z) \cos(\theta(x,z)) \end{cases} \quad (5)$$

creating the Cartesian coordinate data block $D(x', y, z')$ where the data points in the X- and Z-direction have been relocated so that the distortion artifacts in the X-direction are corrected for. Next, in part III of fig.5, a similar procedure is applied to the coordinate points in the second scanning dimension. Similarly to the above, the set of radii $R'(y_c, z')$ is determined experimentally for each central column line $(y = y_c, z')$ of each $B_y$-scan for every z. The polar coordinate equivalents $(R', \varphi)$ originating from the virtual origin of distortion $(y_c, z'_o)$ of the $(y, z')$ coordinate couples are determined as:

$$\forall z' \begin{cases} R'(y_c, z') \quad experimentally\ determined \\ \forall y: \ \varphi(y, z') = tan^{-1}\left(\dfrac{y - y_c}{R'(y_c, z')}\right) \end{cases} \quad (6)$$

where $\varphi(y, z')$ represents the angle between $\vec{R}'(y, z')$ and the central column line $\vec{R}'(y_c, z')$ of the currently processed $B_y$-scan. Next, the same value of $R'(y_c, z')$ is assigned to all coordinate points that have the same z-values, and this process is repeated for all $B_y$-scans. By maintaining the respective angles $\varphi(y, z')$ of all coordinate points, the data block is reshaped so that the distortion effect caused by the second scanning direction is corrected for, as illustrated in part III of fig.5:

$$\forall y, z' \quad \begin{cases} R'(y, z') := R'(y_c, z') \\ \varphi(y, z') = \varphi(y, z') \end{cases} \quad (7)$$

This polar coordinate system can be reconverted to its Cartesian equivalent by applying the following transformation to all $(y, z')$-coordinate couples:

$$\forall y, z' \quad \begin{cases} y' = R'(y_c, z') \sin(\varphi(y, z')) \\ z'' = R'(y_c, z') \cos(\varphi(y, z')) \end{cases} \quad (8)$$

creating the final coordinate location points $(x', y', z'')$ that have been relocated to floating point coordinates so that the distortion artifacts in both scanning directions are now compensated for. As these coordinate point locations are no longer equally spaced on a cubic Cartesian grid, correct visualization of the OCT data volume can only be achieved after performing scattered data interpolation between these floating point coordinates at the predefined Cartesian coordinate grid $(i, j, k) \in \mathbb{N}^3$:

$$D(i, j, k) = I'_{x,y,z} \quad (9)$$

producing new intensity values $I'_{x,y,z}$ that can now correctly be visualized on a 3D Cartesian grid of which the respective voxel sizes have previously been determined in section 2.

**Higher order lens distortion compensation**

When the optical geometry is significantly distorted by the imaging lens L1 (or any other additional non-radial effect), the preceding calibration procedure can easily be expanded to incorporate these 2[nd] order distortion effects by adding an extra term $R^*$ to equations (4) and (7) that represents the non-radial fraction of the distortion effect as perceived by every individual data point:

$$\forall x, z \quad \begin{cases} R(x, z) := R(x_c, z) + R^*(x, z) \\ \theta(x, z) = \theta(x, z) \end{cases} \quad (10)$$

$$\forall y, z' \quad \begin{cases} R'(y, z') := R'(y_c, z') + R^*(y, z') \\ \varphi(y, z') = \varphi(y, z') \end{cases} \quad (11)$$

where $R^*(x, z)$ and $R^*(y, z')$ can be determined after fitting a function of choice through the data points that represent the imaged mirror surface (as shown in fig.4) and calculating the radial differences to the first order circular fit for every point, B-scan per B-scan.

**Application example**

As an application example, an OCT data stack of 12.35 x 10.13 x 2.36 mm³ of a human tympanic membrane was obtained and distortion corrected using the above described calibration procedure. As the shape of this membrane is highly curved in both scanning directions, proper recalibration is mandatory in order to obtain an accurate 3D model of the entire tissue structure. The results of both the corrected and the uncorrected OCT volume were compared quantitatively against a high-resolution topogram, obtained with a fully calibrated Moiré profilometry setup (Buytaert et al. 2008, Dirckx et al. 2010).

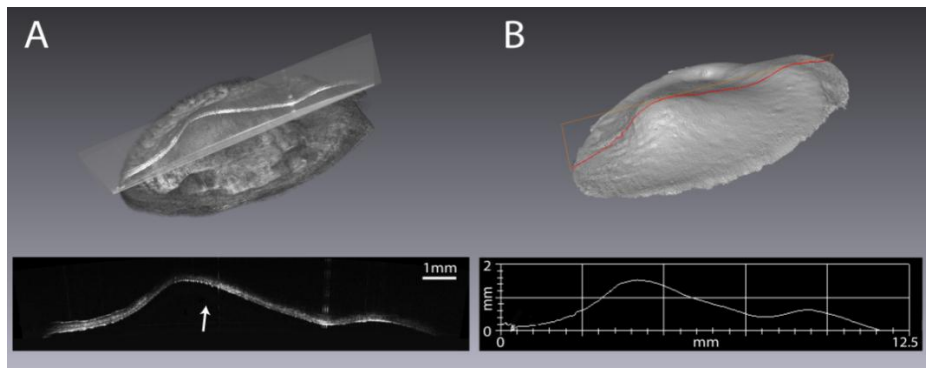

Figure 6: Segmentation of the recorded OCT volume (A) produces a triangulated model (B) of the data stack, from which the object surface shape can easily be extracted. This process converts the OCT greyscale data to a binary surface mesh. A cross-section of the central column data of the measured eardrum is included before segmentation (left) and after segmentation (right).

This well-established optical measurement technique uses fringe interferometry to map height variations with great accuracy. In addition, the employed Moiré system contains a telecentric lens setup to drastically limit the measurement errors Buytaert et al. 2012) to within the 15 μm-range. As Moiré profilometry is restricted to surface mapping only, the acquired OCT volumes were first segmented in specialized visualization software (*Amira* 5.3) to extract the topology of the measured tympanic membrane from the same side at which the Moiré profile was obtained, as directed by the small arrow in fig. 6. After segmentation, the surface meshes of both the uncorrected and the corrected OCT data can be extracted from the triangulated eardrum model. Next, these surface models are aligned to the Moiré model by minimizing the distance between the surfaces, measured at each point of a selectable region of interest on the object surface. The root mean square (RMS) value of these distance calculations is a reliable indicator for the surface alignment in 3 dimensions. As the degrees of freedom in the alignment algorithm were restricted to three translational and three rotational dimensions, no scaling or other affine transformations were allowed, so as to ensure the authentic determination of the surface quality. After alignment, the RMS of the uncorrected model is measured to be RMSu = 45.24 μm, whereas the corrected data stack improves the alignment to RMSc = 13.49 μm.

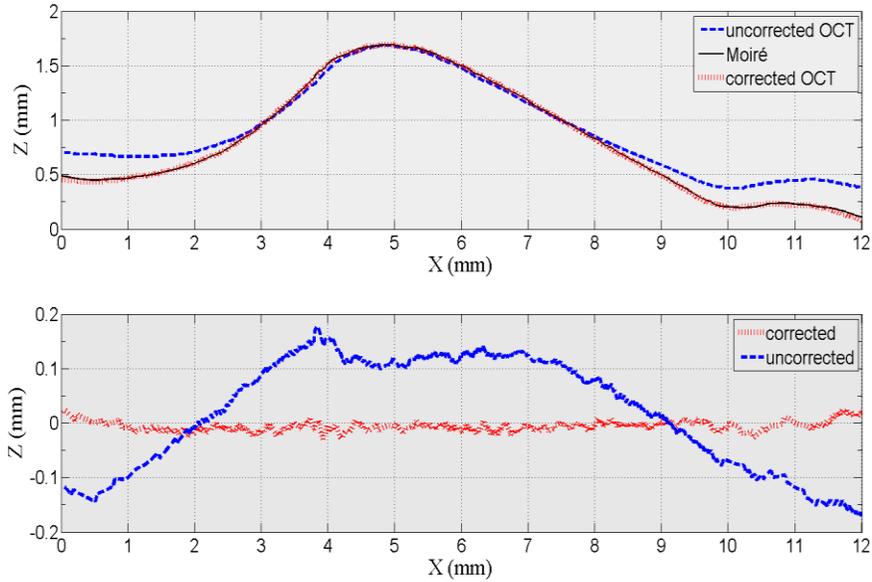

Figure 7: 2D comparison of surface alignment between uncorrected OCT (dashed blue line), corrected OCT (vertically dashed red line) and Moiré (black full line) cross-sections. The bottom graph represents the mathematical difference between the Moiré cut and the OCT cuts.

After three-dimensional alignment, identical surface cuts were made on all three surface models to compare the effect of the calibration algorithm in two dimensions. To most imaging applications, the depth precision of the OCT model is of high importance. Therefore, the *z*-values along the central line in the *X*-direction are plotted in fig. 7. The superior quality of the corrected model over the uncorrected model is immediately noted by examining the depth variation of the membrane near the edges of the scanning field. To highlight the deviation from the Moiré surface (black full line), a difference plot is included in the bottom graph of fig.7 for both the uncorrected (blue dashed line) and the corrected (red vertically dashed line) model. Near the edges, the uncorrected surface model is misaligned to the Moiré model by almost 200 μm, whereas the corrected model misalignment stays within the 20 μm-range throughout the full length of the B-scan.

**Real-time implementation**

Without performing the refractive index correction, described earlier in step 1, full 3D reconstruction of the observed volume can be achieved without prior knowledge of the internal sample structure. As we have demonstrated that a succession of coordinate transformations suffices to accurately remap the 3D data acquired with a standard OCT setup, the technique described by Van der Jeught et al. (2012) can be used to create a look-up table of floating point coordinates at which the recorded OCT data blocks need to be interpolated in order to remove all geometric distortion artifacts. For the employed OCT setup, this look-up table needs to be pre-calculated only once, after which it can be used to correct all incoming data cubes. The required interpolation operations can be implemented very effectively on dedicated memory regions in standard graphics processing units. To assess the processing time delays caused by the proposed technique, we have benchmarked our GPU-based software for different sample sizes, using the NVIDIA CUDA profiler. The results that we achieved with our system, containing a 3.33 GHz Intel i5 quad-core CPU and a GTX570 graphics card with 480 CUDA processing cores and 1280MB of on-board memory, are included in Table 1.

| Volume size (pixels) | Volume size (MB) | Memory transfer (ms) | 3D Interpolation (ms) |
|---|---|---|---|
| 256×256×256 | 67.1 | 24.2 | 0.9 |
| 512×512×256 | 268.4 | 102.2 | 3.6 |
| 512×512×512 | 536.8 | 194.7 | 7.1 |
| 768×768×512 | 1207.9 | 436.8 | 10.4 |

Table 1: Benchmarks of 3D interpolation operations when implemented on a graphics processing unit of a standard PC. Different sample sizes have been tested and processing times for both memory transfers between host and device, and the isolated interpolation functions are presented.

Several recent reports have been published on GPU-based processing setups that capture, process and render several 3D OCT data blocks per second, depending on the observed volume size (Zhang et al. 2010, Young et al. 2011, Huang et al. 2012), without any of them performing geometric distortion correction in real-time. These processing schemes all have in common that the incoming OCT data needs to be transferred to GPU memory before any processing on the device can be done. For such setups, column 4 of Table 1 suggests that the additional cost of performing the proposed interpolation operations in 3D can be kept low, as the OCT volume is already present in the GPU memory and no extra data transfers need to be made. Notice how the processing times of the interpolation calculations rise linearly with exponential increase in data size, illustrating the parallel design of modern GPU processing core architecture. Notice also how the memory transfer times between host and device, included in column 3 of Table 1, significantly outweigh the actual processing times of the interpolation operations on the device. Even for the largest possible volume size that can be transferred to the on-board memory of the employed GPU, the interpolation calculations impose little or no lag on the rendering rate, while the corresponding memory transfer times alone limit the maximum volume rate to about 2 volumes/s. This behavior confirms one of the conclusions drawn by Li et al. (2011) on recent developments of GPUs used in optical coherence tomography setups; namely that the maximum achievable line rate in OCT systems is currently dictated by the memory transfer times and not by the processing times.

## DISCUSSION

It can be noticed that the corrected OCT surface approximates the Moiré surface well almost everywhere except near the edges of the membrane. The same effect occurs in the uncorrected model, where the misalignment is more pronounced and the net difference has an opposite sign. This opposite behavior indicates that the radial distortion approach we have used in the recalibration procedure slightly overcompensates for the actual probing beam path, due to the higher order distortion effects introduced by the imaging lens L1. Nevertheless, for the employed setup, the radial distortion correction approach is sufficiently accurate to correct for the entire OCT data volume (in air). As the $RMSc = 13.49$ µm of the corrected data shape is less than the intrinsic margin of error (<15µm) of the aberration-free optical Moiré setup, no higher order corrections were applied. Currently, accurate sample structure information is typically gathered in post-processing by applying ray-tracing software to the processed OCT images. By calculating the optical pathway for every light ray individually, ray-tracing is able to accurately pinpoint the origin location of reflection within the observed sample and thus relocate the pixels in the resulting data block accordingly. These calculations come at a great computational cost on the one hand, and require prior information on the sample microstructure to account for refraction artifacts on the other hand, making it unfeasible to be applied in real-time.  The approach presented in this chapter however, is capable of straightforwardly compensating for all geometric distortion artifacts induced by the scanning setup, including fan distortion and any higher order lens effects that may affect the acquired data volume. The major advantage of the proposed technique with respect to the previously mentioned ray-tracing based schemes is its ability to be applied at near-instant speeds using an off-the-shelf GPU board. Despite the fact that the data transfers between host and device currently limit the scalability of real-time OCT performance, most real-time volumetric OCT processing algorithms employ GPUs anyway to process the incoming data in real-time. We have shown that the added time delay for distortion correcting incoming data blocks in 3D is minimal if the data is already transferred to device memory. While the presented correction procedure in no way claims to replace the additional refraction error corrections that can be calculated in post-processing, it does provide the system operator with more accurate shape information on the observed sample when applied in real-time. Since we have demonstrated that following a simple calibration procedure suffices to accurately remove all geometric distortion artifacts for a given setup, both pre-and post-processing methods can in fact coexist.

## CONCLUSION

A three-dimensional recalibration procedure was described that effectively compensates for geometric distortion artifacts as they occur in setups where large scanning angles are required to obtain a large field of view. The predominantly radial nature of this type of distortion, often found in applied optical coherence tomography, was demonstrated for an OCT volume of 12.35 x 10.13 x 2.36 mm³. The quality of the proposed 3D recalibration scheme was confirmed by aligning the surface mesh of a highly curved tympanic membrane that was obtained with the described OCT setup to the surface model of the membrane, obtained using optical Moiré profilometry. After application of the correction method, the two surfaces coincide within the <15µm margin of error of the highly accurate Moiré setup. By implementing the proposed interpolation operations on the graphics card of a standard PC, we have demonstrated that the additional processing time required for distortion correcting large 3D OCT data sets is limited to nearly 10 ms, which enables its real-time implementation. In contrast to previously reported distortion correction methods, this procedure allows real-time rendering of distortion-free surface information. This provides the user with considerable diagnostic advantages in any area where real-time monitoring of surface shape behavior is requisite.

## ACKNOWLEDGMENTS

The work presented in this chapter was funded by the Research Foundation – Flanders (FWO). A. Bradu and A. Podoleanu acknowledge the support of the ERC grant COGATIMABIO, 249889.

## REFERENCES

Amaechi, B. T., Higham, S. M., Podoleanu, A., Rogers, J. A., & Jackson, D. A. 2001. Use of optical coherence tomography for assessment of dental caries: quantitative procedure. *Journal of Oral Rehabilitation*, 28(12), pp. 1092-1093.

Bechmann, M., Thiel, M. J., Neubauer, A. S., Ullrich, S., Ludwig, K., Kenyon, K. R., & Ulbig, M. W. 2001. Central corneal thickness measurement with a retinal optical coherence tomography device versus standard ultrasonic pachymetry. *Cornea*, 20(1), pp. 50-54.

Bibas, A. G.,et al. 2004. 3-D optical coherence tomography of the laryngeal mucosa. *Clinical Otolaryngology & Allied Sciences*, 29(6), 713-720.

Buytaert, J. A., & Dirckx, J. J. 2008. Moiré profilometry using liquid crystals for projection and demodulation. *Optics Express*, 16(1), pp. 179-193.

Buytaert, J.A.N., Ribbens, B., Vanlanduit, S. & Dirckx, J.J.J., 2012. Aberration-free moiré profilometry—Analysis, simulation and implementation of the optimal setup geometry. *Optics and Lasers in Engineering*, 50, pp. 1119–1129.

Choma, M. A., Sarunic, M. V., Yang, C., & Izatt, J. A. 2003. Sensitivity advantage of swept source and Fourier domain optical coherence tomography. *Optics Express*, 11(18), pp. 2183-2189.

Dirckx, J. J., Buytaert, J. A., & Van der Jeught, S. 2010. Implementation of phase-shifting moiré profilometry on a low-cost commercial data projector. *Optics and Lasers in Engineering*, 48(2), pp. 244-250.

Drexler, W. & Fujimoto, J.G. 2008. Optical coherence tomography: technology and applications, Springer, Chapter 13.7 Image Representation and Coloring, pp. 417-19.

Fercher, A. F., Drexler, W., Hitzenberger, C. K., & Lasser, T. 2003. Optical coherence tomography-principles and applications. *Reports on progress in physics*, 66(2), pp. 239.

Huang, D., et al. 1993. *Optical coherence tomography* (Doctoral dissertation, Massachusetts Institute of Technology, Whitaker College of Health Sciences and Technology).

Huang, Y., Liu, X., & Kang, J. U. 2012. Real-time 3D and 4D Fourier domain Doppler optical coherence tomography based on dual graphics processing units. *Biomedical Optics Express*, 3(9), pp. 2162-2174.

Kerbage, C., Lim, H., Sun, W., Mujat, M., & de Boer, J. F. 2007. Large depth-high resolution full 3D imaging of the anterior segments of the eye using high speed optical frequency domain imaging. *Optics express*, 15(12), pp. 7117-7125.

Kim, E., Ehrmann, K., Uhlhorn, S., Borja, D., & Parel, J. M. 2009. Automated analysis of OCT images of the crystalline lens. *In Proc. SPIE*, Vol. 7163.

Kraus, M. F., et al.. Motion correction in optical coherence tomography volumes on a per A-scan basis using orthogonal scan patterns. *Biomedical Optics Express*, 3(6), pp. 1182-1199.

Leitgeb, R., Hitzenberger, C. K., & Fercher, A. F. 2003. Performance of fourier domain vs. time domain optical coherence tomography. *Opt. Express*, 11(8), pp. 889-894.

Li, J., Bloch, P., Xu, J., Sarunic, M. V., & Shannon, L. 2011. Performance and scalability of Fourier domain optical coherence tomography acceleration using graphics processing units. *Applied optics*, 50(13), pp. 1832-1838.

McNabb, R. P., LaRocca, F., Farsiu, S., Kuo, A. N., & Izatt, J. A. 2012. Distributed scanning volumetric SDOCT for motion corrected corneal biometry. *Biomedical Optics Express*, 3(9), pp. 2050-2065.

Miura, M., et al. 2007. Three-dimensional optical coherence tomography of granular corneal dystrophy. *Cornea*, 26(3), pp. 373-374.

Nassif, N. A., et al. 2004. In vivo high-resolution video-rate spectral-domain optical coherence tomography of the human retina and optic nerve. *Opt. Express*, 12(3), pp. 367-376.

Ortiz, S., Siedlecki, D., Grulkowski, I., Remon, L., Pascual, D., Wojtkowski, M., & Marcos, S. 2010. Optical distortion correction in optical coherence tomography for quantitative ocular anterior segment by three-dimensional imaging. *Optics express*, 18(3), pp. 2782-2796.

Ortiz, S., Siedlecki, D., Remon, L., & Marcos, S. 2009. Optical coherence tomography for quantitative surface topography. *Applied optics*, 48(35), pp. 6708-6715.

Radhakrishnan, S., Rollins, A. M., Roth, J. E., Yazdanfar, S., Westphal, V., Bardenstein, D. S., & Izatt, J. A. 2001. Real-time optical coherence tomography of the anterior segment at 1310 nm. *Archives of Ophthalmology*, 119(8), pp. 1179.

Raffel, O. C., Akasaka, T., & Jang, I. K. 2008. Cardiac optical coherence tomography. *Heart*, 94(9), pp. 1200-1210.

Schmitt, J. M. 1999. Optical coherence tomography (OCT): a review. *Selected Topics in Quantum Electronics*, IEEE Journal of, 5(4), 1205-1215.

Sepehr, A., Djalilian, H. R., Chang, J. E., Chen, Z., & Wong, B. J. 2008. Optical coherence tomography of the cochlea in the porcine model. *The Laryngoscope*, 118(8), pp. 1449-1451.

Trifanov, I., et al. 2011. Combined neodymium–ytterbium-doped ASE fiber-optic Source for optical coherence tomography applications. *Photonics Technology Letters, IEEE*, 23(1), pp. 21-23.

Tripathi, R., Nassif, N., Nelson, J. S., Park, B. H., & de Boer, J. F. 2002. Spectral shaping for non-Gaussian source spectra in optical coherence tomography. *Optics Letters*, 27(6), pp. 406-408.

Van der Jeught, S., Bradu, A., & Podoleanu, A. G. 2010. Real-time resampling in Fourier domain optical coherence tomography using a graphics processing unit. *Journal of Biomedical Optics*, 15(3), pp. 030511.

Von Unge, M., Decraemer, W. F., Bagger-Sjöbäck, D., & Dirckx, J. J. 1993. Displacement of the gerbil tympanic membrane under static pressure variations measured with a real-time differential moiré interferometer. *Hearing research*, 70(2), pp. 229-242.

Westphal, V., Rollins, A., Radhakrishnan, S., & Izatt, J. 2002. Correction of geometric and refractive image distortions in optical coherence tomography applying Fermat's principle. *Optics Express*, 10(9), pp. 397-404.

Wojtkowski, M., Srinivasan, V., Fujimoto, J. G., Ko, T., Schuman, J. S., Kowalczyk, A., & Duker, J. S. 2005. Three-dimensional retinal imaging with high-speed ultrahigh-resolution optical coherence tomography. *Ophthalmology*, 112(10), pp. 1734-1746.

Yasuno, Y., et al. 2005. Three-dimensional and high-speed swept-source optical coherence tomography for in vivo investigation of human anterior eye segments. *Optics Express*, 13(26), pp. 10652-10664.

Young, M., Lebed, E., Jian, Y., Mackenzie, P. J., Beg, M. F., & Sarunic, M. V. 2011. Real-time high-speed volumetric imaging using compressive sampling optical coherence tomography. *Biomedical optics express*, 2(9), pp. 2690-2697.

Zhang, K., & Kang, J. U. 2010. Real-time 4D signal processing and visualization using graphics processing unit on a regular nonlinear-k Fourier-domain OCT system. *Optics express*, 18(11), pp. 11772-11784.

Zhao, M., Kuo, A. N., & Izatt, J. A. 2010. 3D refraction correction and extraction of clinical parameters from spectral domain optical coherence tomography of the cornea. *Optics Express*, 18(9), pp. 8923-8936.